\documentclass{aa}
\usepackage{graphics}
\begin{document}
\thesaurus{03
              (12.03.3;  
               12.12.1;  
               11.03.1;
               11.04.1}

\title{Spectroscopic confirmation of clusters from the
ESO imaging survey \thanks{Based on observations collected at the European
Southern Observatory (La Silla, Chile), Proposal ID: 62.O-0601}}


\author{M.~Ramella \inst{1}
   \and A.~Biviano \inst{1}
   \and W.~Boschin \inst{1}
   \and S.~Bardelli \inst{2}
   \and M.~Scodeggio \inst{3,4}
   \and S.~Borgani \inst{5,6}
   \and C.~Benoist \inst{4,7}
   \and L.~da~Costa \inst{4}
   \and M.~Girardi \inst{8}
   \and M.~Nonino \inst{1,4}
   \and L.F.~Olsen \inst{9}
          }

\offprints{M. Ramella, ramella@oat.ts.astro.it}

\institute{Osservatorio Astronomico di Trieste,
via G.B. Tiepolo 11, I-34131 Trieste, Italy
\and
Osservatorio Astronomico di Bologna,
via Ranzani 1, I-40127 Bologna, Italy 
\and
Istituto di Fisica Cosmica ``G. Occhialini'',
via Bassini 15, I-20133 Milano, Italy
\and 
European Southern Observatory, Karl-Schwarzschild-Strasse 2,
D-85748 Garching bei M\"unchen, Germany
\and
INFN, Sezione di Trieste, c/o Dipartimento di Astronomia, Universit\`a
degli Studi di Trieste, via G.B. Tiepolo 11, I-34131 Trieste, Italy
\and
INFN, Sezione di Perugia, c/o Dipartimento di Fisica dell'Universit\`a,
via A. Pascoli, I-06123, Perugia, Italy
\and
DAEC, Observatoire de Paris-Meudon, 5 Pl. J. Janssen,
F-92195 Meudon Cedex, France
\and
Dipartimento di Astronomia, Universit\`a degli Studi di Trieste,
via G.B. Tiepolo 11, I-34131 Trieste, Italy
\and
Copenhagen University Observatory, Juliane Maries Vej 30, DK-2100,
Copenhagen, Denmark
}

\date{Received March 23, 2000; accepted May 9, 2000}

\authorrunning{Ramella et al.}
\titlerunning{Spectroscopic confirmation of EIS clusters}

\maketitle

\begin{abstract} 

We measure redshifts for 67
galaxies in the field of six cluster candidates from the ESO Imaging
Survey (EIS). The cluster candidates are selected in the EIS patches
C and D among those with estimated mean redshifts $0.5 \leq z \leq 0.7$.
The observations were made with EFOSC2 at the 3.6m ESO telescope.

In the six candidate cluster fields, we identify 19 possible sets of 2
to 7 galaxies in redshift space.  In order to establish which of the
19 sets are likely to correspond to real dense systems we compare our
counts with those expected from a uniform distribution of galaxies
with given luminosity function. In order to take into account the
effect of the Large Scale Structure, we modulate the probability
computed from the luminosity function with random samplings of the
Canada-France Redshift Survey.

We find that four out of six candidate EIS clusters are likely to
correspond to real systems in redshift space ($> 95$~\% confidence
level).  Two of these systems have mean redshift in agreement with the 
redshift estimate given by the matched filter algorithm ($\Delta z
= \pm 0.1$). The other two systems have significantly lower
redshifts.

We discuss the implications of our results in the context of our ongoing
research projects aimed at defining high-redshift optically-selected cluster
samples.

\end{abstract}

\section{Introduction}\label{s-int}
Clusters of galaxies are the largest virialized structures observed in
the Universe. Since they arise from exceptionally high peaks of the
primordial fluctuation density field, their properties are highly
sensitive to the nature of such cosmic fluctuations. Therefore, the
mass function of both local (e.g. White et al. \cite{whi93}; Girardi
et al. \cite{gir98}) and distant clusters (e.g. Oukbir \& Blanchard
\cite{ouk92}; Carlberg et al. \cite{car97}; Eke et al. \cite{eke98};
Borgani et al. \cite{bor99}) is a powerful tool to constrain
cosmological models for the formation and evolution of cosmic
structures.  Moreover, clusters are useful laboratories for testing
models of galaxy evolution. While early-type galaxies only show
evidence for passive evolution (e.g. Stanford et al.  \cite{sta98}),
the fraction of blue galaxies increases significantly with redshift
(Butcher \& Oemler \cite{but78}), at least up to $z \sim 0.5$, and the
fraction of S0's decreases (Dressler et al. \cite{dre99}).  It is
therefore essential to have reliable cluster catalogues over the
largest possible redshift range.

Most distant clusters, at $z \geq 0.5$, have so far been identified
through optical follow-ups of X-ray selected clusters (see, e.g. Gioia
et al. \cite{gio90} and Rosati et al. \cite{ros00} for a recent
review), or by looking at the environment of high-redshift radio
galaxies (e.g. Smail \& Dickinson \cite{sma95}; Deltorn et
al. \cite{del97}).

In the optical, clusters at $z \simeq 0.5$ and beyond started to be
classified in the 80's (Gunn et al. \cite{gun86}). In the 90's a large
catalogue of objectively selected distant clusters, identified in the
optical, became available (Postman et al. \cite{pos96}). These last
clusters are identified using a matched-filter algorithm using both
positional and photometric data. In brief, this algorithm filters a
galaxy catalogue to remove fluctuations in the projected distribution
of galaxies that are not likely to be galaxy clusters. For this
purpose, the filter is built around parametrizations of the spatial
distribution and luminosity function of cluster galaxies. This
algorithm also provides an estimate of the redshift for each candidate
cluster (hereafter we refer to the matched-filter estimated redshift
as $z_{mf}$). Currently, $\simeq 30$ PDCS clusters have been confirmed
spectroscopically, most of them at $z < 0.5$ (Holden et
al. \cite{hol99a}, \cite{hol99b}; Oke et al. \cite{oke98}).

Recently, Olsen et al. (\cite{ols99a}, \cite{ols99b}) and Scodeggio et
al. (\cite{sco99}) have presented a catalogue of 302 cluster
candidates from the $I$-band images of the ESO Imaging Survey (EIS,
see Renzini \& da Costa \cite{ren97}).  Clusters are identified in
two dimensions (hereafter, 2-d) using the matched filter algorithm of
Postman et al. (\cite{pos96}; see Olsen et al. \cite{ols99a}). The
estimated redshifts for EIS clusters span the range $0.2 \leq z_{mf}
\leq 1.3$, with a median redshift $z_{mf}=0.5$. 

Several EIS cluster candidates have been confirmed so far, most at
$z < 0.5$, either from the existence of the red sequence of
cluster ellipticals/S0's in colour-magnitude diagrams (Olsen et
al. \cite{ols99b}), or from a combination of photometric and
spectroscopic data (da Costa et al. \cite{dac99}).

The EIS cluster catalogue is the largest optically selected cluster
sample currently available in the Southern Hemisphere to this depth.
This catalogue constitutes an obvious reference for follow-up
observations at the ESO VLT aimed at determining the structure and
dynamics of distant clusters, as well as the spectroscopic properties
of their member galaxies.  Unfortunately, little is currently known on
the performance of the matched filter algorithm in detecting real
clusters at $z \geq 0.5$. As we already pointed out, most confirmed
PDCS and EIS clusters have redshifts $z < 0.5$.  Therefore, to point
blindly at EIS cluster candidates would make for an inefficient use of
VLT time, because we expect several of these candidate clusters not to
be real, in particular at $z_{mf} \geq 0.5$.

The aim of our investigation is twofold: we want to confirm as many
EIS clusters as possible, in order to build a reliable sample of
distant clusters with well determined redshift, and, at the same time,
evaluate the performance of the matched filter algorithm in the
detection of high-redshift clusters. In order to achieve this purpose,
we use two independent methods: (1) multi-object spectroscopic
observations of EIS cluster candidates in the redshift range $0.5 \leq
z_{mf} \leq 0.7$, and (2) the detection of the colour-magnitude
sequences traced by early-type galaxies through multi-colour optical
and near-IR photometry of the most distant EIS cluster candidates
(Scodeggio et al., in preparation). 

In this paper we report the first results of the spectroscopic
investigations of 6 EIS clusters. We are able to confirm the existence
of significant concentrations in redshift space in correspondence of four of
the six EIS fields targeted.  For two of these confirmed clusters, the
spectroscopic mean redshift agrees with the matched-filter estimate to
within $\Delta z = \pm 0.1$.

In Sect.~\ref{s-obs} we describe our spectroscopic observations,
data reduction, and give the new galaxy redshifts.  In
Sect.~\ref{s-zsys} we analyse the data, and define sets of galaxies
in redshift space. We also discuss the concordance of the mean
redshifts of these sets with the matched-filter estimates of the
cluster mean redshifts.  We then make a likelihood analysis of the
reality of the galaxy sets, and flag four of them as reliable at $>
95$~\% confidence level (Sect.~\ref{s-likely}).  Finally, we discuss
our results and give our conclusions in Sect.~\ref{s-con}.

We use H$_0=$ h$_{75}$~75~km~s$^{-1}$~Mpc$^{-1}$, $\Omega_0=0.2$ and
$\Omega_{\Lambda}=0$ throughout this paper, unless otherwise
stated.

\section{Observations and data reduction}\label{s-obs}
We select our targets among the candidate EIS clusters in patches C
and D, with estimated redshift $0.5 \leq z_{mf} \leq 0.7$ (Scodeggio
et al. \cite{sco99}).  We do not apply any additional criterion for
the selection of our cluster candidates. The size of our sample is
one tenth of the total of 36$+$28 EIS clusters in patches C and D
within the above-mentioned redshift range.

Our targets are listed in Table~\ref{t-sc}. In Column (1) we list the
cluster candidate identification name, in Column (2) and (3) the right
ascension and declination (J2000), in Column (4) the cluster richness
(see Olsen et al. \cite{ols99a}), and in Column (5) the matched-filter
redshift estimate, $z_{mf}$ (see Scodeggio et al. \cite{sco99}). 
In Column (6) we list the number of galaxies targeted for multi-slit
spectroscopy in each cluster field, and in Column (7) the number of
successful redshift estimates.

\begin{table*}
\caption[ ]{Cluster candidates selected for observation}
\label{t-sc}
\begin{tabular}{rrrrrrr}
\hline
Id. & $\alpha_{J2000}$ & $\delta_{J2000}$ &
$\Lambda_{cl}$ & $z_{mf}$ & n$_{slits}$ & n$_{z}$  \\
\hline
EIS0533-2353 & 05:33:36.5 & -23:53:52.9 & 72.0 & 0.7 & 16 & 12 \\
EIS0540-2418 & 05:40:08.5 & -24:18:19.3 & 83.8 & 0.6 & 20 & 14 \\
EIS0950-2154 & 09:50:47.9 & -21:54:38.3 & 62.4 & 0.5 & 18 & 11 \\
EIS0951-2047 & 09:51:32.5 & -20:47:08.3 & 65.1 & 0.5 & 15 & 11 \\
EIS0955-2113 & 09:55:32.3 & -21:13:55.3 & 68.2 & 0.6 & 17 &  7 \\
EIS0956-2009 & 09:56:28.6 & -20:09:27.4 & 58.2 & 0.5 & 16 & 13 \\
\hline
\end{tabular}
\end{table*}

The observations were carried out at the 3.6~m ESO telescope at La
Silla, Chile, during two nights in February 1999.  The weather
conditions were good, with seeing slightly above 1'', in partial
moonlight.  With a total useful observing time of 16 hours over the
two nights, we were able to obtain $3 \times 45$~min exposures for
each cluster. 

We observed with EFOSC2 in Multi-Object Spectroscopy (MOS)
mode. EFOSC2 was equipped with a Loral CCD of 2048$\times$2048 with
15~$\mu$m pixels, allowing for an unvignetted field-of-view of
3.8'$\times$5.5'.  We used Grism \# 1, giving a spectral range
3185--10940 \AA, and a dispersion of 6.3 \AA/pixel. On the MOS masks
our slits were 1.2'' wide.

\begin{figure*}
\resizebox{\hsize}{!}{\includegraphics{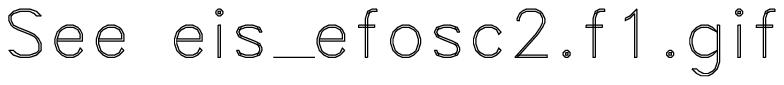}}
\caption{An $I_c$-band image of the EIS0533-2353 cluster candidate.
The circles indicate those galaxies for which we obtained redshifts.
The field is 6.4'$\times$4.3', centered at $\alpha=$05:33:36.5, $\delta=-$23:53:53
(J2000). North is up, East is to the left.}
\label{f-m1}
\end{figure*}

\begin{figure*}
\resizebox{\hsize}{!}{\includegraphics{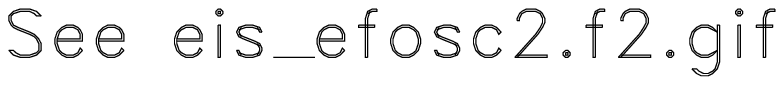}}
\caption{Same as Fig.~\ref{f-m1}, for the EIS0540-2418 cluster candidate.
The field is centered at $\alpha=$05:40:08.5, $\delta=-$24:18:19 (J2000).}
\label{f-m2}
\end{figure*}

\begin{figure*}
\resizebox{\hsize}{!}{\includegraphics{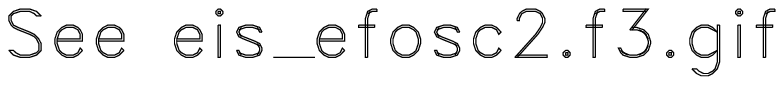}}
\caption{Same as Fig.~\ref{f-m1}, for the EIS0950-2154 cluster
candidate.  Large circles indicate galaxies identified as members of a
significant overdensity in redshift space.  The field is centered at
$\alpha=$09:50:48.6, $\delta=-$21:55:15 (J2000).}
\label{f-m3}
\end{figure*}

\begin{figure*}
\resizebox{\hsize}{!}{\includegraphics{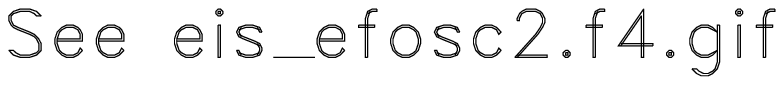}}
\caption{Same as Fig.~\ref{f-m1}, for the EIS0951-2047 cluster candidate.
Large circles indicate galaxies identified as members of a 
significant overdensity in redshift space.
The field is centered at $\alpha=$09:51:31.8, $\delta=-$20:46:57 (J2000).}
\label{f-m4}
\end{figure*}

\begin{figure*}
\resizebox{\hsize}{!}{\includegraphics{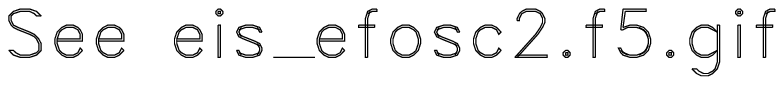}}
\caption{Same as Fig.~\ref{f-m1}, for the EIS0955-2113 cluster candidate.
Large circles are galaxies members of a significant overdensity 
in redshift space.
The field is centered at $\alpha=$09:55:32.1, $\delta=-$21:13:55 (J2000).}
\label{f-m5}
\end{figure*}

\begin{figure*}
\resizebox{\hsize}{!}{\includegraphics{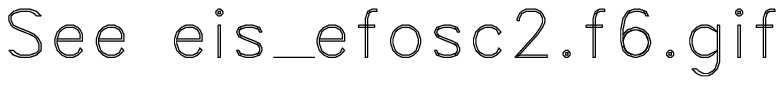}}
\caption{Same as Fig.~\ref{f-m1}, for the EIS0956-2009 cluster candidate.
Large circles indicate galaxies identified as members of a 
significant overdensity in redshift space.
The field is centered at $\alpha=$09:56:28.4, $\delta=-$20:09:30 (J2000).}
\label{f-m6}
\end{figure*}

We obtained spectra for 102 objects in the magnitude range $17.0 \leq
m_I \leq 21.3$, where $m_I$ is the apparent magnitude in the $I_c$
band (Nonino et al. \cite{non99}).  In Fig.~\ref{f-m1}--\ref{f-m6} we
show $I_c$-band images of the six EIS candidate clusters. Small
circles mark galaxies with redshift, large circles mark galaxies
belonging to significant overdensities in redshift space (see
Sect.~\ref{s-likely}).


We reduce the data with standard IRAF\footnote{IRAF is distributed by
the National Optical Astronomy Observatories, which is operated by
AURA Inc. under contract with NSF} packages.  We determine redshifts
using the task XCSAO that implements the cross-correlation technique
of Tonry \& Davis (\cite{ton79}). We use several real and synthetic
templates for the cross-correlation. We use emission lines, where
present in the spectrum of the object, to determine the redshift with
the task EMSAO.  We examine visually all
spectra, by overplotting the positions of the major spectral features
redshifted at the redshift(s) determined by the automatic techniques
described above. We employ particular care in flagging those features
that could be contaminated by night-sky lines.

\begin{figure}
\resizebox{\hsize}{!}{\includegraphics{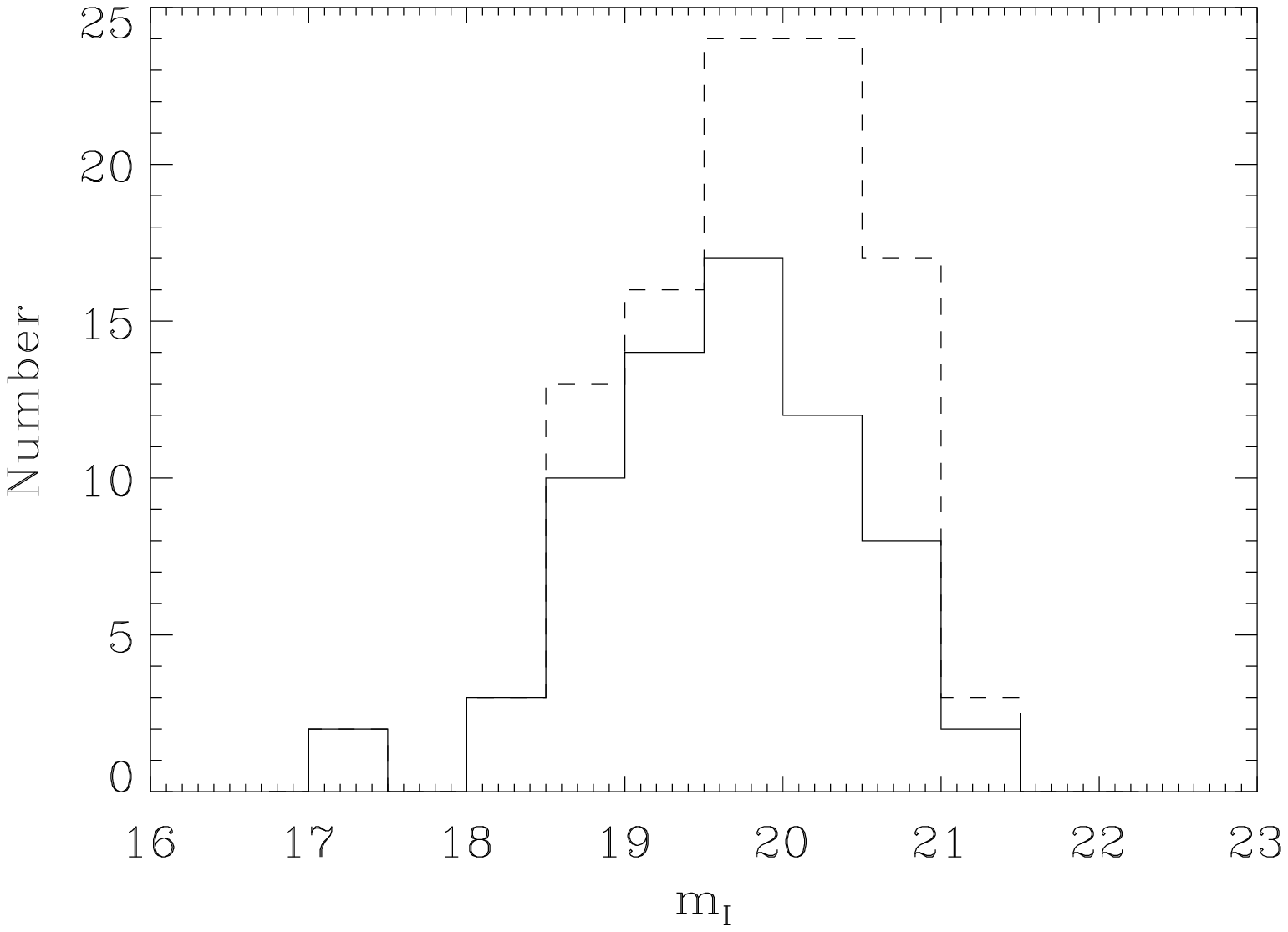}}
\caption{The $I_c$-band magnitude distribution of the 102 targets for
which we attempted spectroscopy (dashed-line histogram), and of the 67
galaxies (plus a QSO) for which we obtained a redshift (solid-line histogram).}
\label{f-md}
\end{figure}

In total, we determine 67 galaxy redshifts, from a minimum of $z=0.09$
to a maximum of $z=0.79$, with an average $\overline{z}=0.380$. One of
our objects turns out to be a QSO at $z=3.2$. We do not consider this
object in our analysis. An internal
estimate of the typical redshift uncertainty is $\delta z \sim 0.001$.
The success-rate is magnitude-dependent, as can be seen in Fig.~\ref{f-md}:
it is 85~\% for $m_I \leq 19.5$ and decreases to 57~\% for fainter galaxies.

We list in Table~\ref{t-lg} the galaxies with measured redshift. In Column (1)
we list the name of the EIS candidate cluster field, in Column (2) a
galaxy identification number, in Column (3) and (4) the (J2000) right
ascension and the declination of the galaxy, in Column (5) the $I_c$
magnitude, in Column (6) the redshift, and in Column (7) the galaxy set
to which the galaxy is assigned. These galaxy sets
are defined in Sect.~\ref{s-zsys} and listed in Table~\ref{t-cs}.

\begin{table*}
\caption[ ]{Galaxies in the six EIS cluster fields}
\label{t-lg}
\begin{tabular}{rrrrrll}
\hline
Field id. & Galaxy id. & $\alpha_{J2000}$ & $\delta_{J2000}$ & $m_I$ & 
$z$ & galaxy set \\
\hline
EIS0533-2353 & 401117 & 05:33:31.16 & -23:51:52.9 &   19.40 &  0.311 & 1b\\
& 391647 & 05:33:31.56 & -23:53:27.6 &   19.04 &  0.332 & 1c\\
& 395828 & 05:33:32.05 & -23:52:46.9 &   20.25 &  0.335 & 1c\\
& 394374 & 05:33:33.78 & -23:52:55.3 &   19.72 &  3.2   & \\
& 391769 & 05:33:33.86 & -23:53:19.3 &   20.25 &  0.274 & 1a\\
& 391224 & 05:33:34.77 & -23:53:27.2 &   20.06 &  0.497 & 1d\\
& 391895 & 05:33:35.18 & -23:53:29.6 &   18.46 &  0.305 & 1b\\
& 387448 & 05:33:35.26 & -23:54:00.6 &   20.52 &  0.509 & 1d\\
& 391850 & 05:33:38.24 & -23:53:19.6 &   20.73 &  0.206 & \\
& 386188 & 05:33:38.58 & -23:54:10.1 &   20.97 &  0.176 & \\
& 391395 & 05:33:42.00 & -23:53:26.4 &   20.70 &  0.42  & \\
& 378902 & 05:33:42.75 & -23:55:21.1 &   19.19 &  0.277 & 1a\\
\hline  		   		    	        	 
EIS0540-2418 & 211948 & 05:40:00.44 & -24:19:11.3 &   19.75 &  0.528 & 2c\\
& 214076 & 05:40:01.47 & -24:18:53.7 &   19.57 &  0.476 & 2b\\
& 215860 & 05:40:05.15 & -24:18:27.4 &   21.18 &  0.467 & 2b\\
& 217648 & 05:40:05.42 & -24:18:15.0 &   20.14 &  0.603 & \\
& 219687 & 05:40:06.33 & -24:17:51.4 &   21.09 &  0.790 & \\
& 217861 & 05:40:08.82 & -24:18:18.6 &   18.59 &  0.534 & 2c\\
& 228555 & 05:40:09.28 & -24:16:37.8 &   19.93 &  0.289 & \\
& 216666 & 05:40:12.93 & -24:18:35.8 &   18.72 &  0.437 & 2a\\
& 231275 & 05:40:12.95 & -24:16:27.5 &   19.94 &  0.087 & \\
& 220131 & 05:40:13.66 & -24:18:00.9 &   20.18 &  0.698 & 2d\\
& 220959 & 05:40:14.98 & -24:17:40.9 &   20.80 &  0.710 & 2d\\
& 222317 & 05:40:15.45 & -24:17:30.4 &   19.77 &  0.442 & 2a\\
& 225021 & 05:40:16.22 & -24:17:10.0 &   20.37 &  0.519 & 2c\\
& 223115 & 05:40:18.12 & -24:17:23.3 &   20.59 &  0.696 & 2d\\
\hline  		   		    	        	 
EIS0950-2154 &  56524 & 09:50:41.48 & -21:53:58.6 &   19.50 &  0.215 & 3b\\
&  51381 & 09:50:44.84 & -21:54:44.2 &   19.35 &  0.285 & \\
&  44805 & 09:50:45.40 & -21:55:48.3 &   19.25 &  0.126 & 3a\\
&  53238 & 09:50:47.12 & -21:54:22.6 &   19.86 &  0.214 & 3b\\
&  53078 & 09:50:47.67 & -21:54:33.1 &   19.08 &  0.486 & \\
&  50696 & 09:50:49.42 & -21:54:42.5 &   20.27 &  0.456 & \\
&  46272 & 09:50:51.12 & -21:55:30.1 &   17.39 &  0.129 & 3a\\
&  45102 & 09:50:52.24 & -21:55:39.2 &   18.64 &  0.240 & 3c\\
&  45447 & 09:50:52.83 & -21:55:32.3 &   19.00 &  0.129 & 3a\\
&  53713 & 09:50:55.24 & -21:54:16.2 &   20.45 &  0.622 & \\
&  35640 & 09:50:56.39 & -21:56:54.7 &   19.59 &  0.242 & 3c\\
\hline  		   		    	        	 
EIS0951-2047 & 504012 & 09:51:23.60 & -20:47:30.4 &   18.46 &  0.281 & \\
& 494234 & 09:51:24.75 & -20:48:33.3 &   19.09 &  0.241 & 4c\\
& 512064 & 09:51:30.08 & -20:46:09.6 &   19.39 &  0.234 & 4c\\
& 505407 & 09:51:30.46 & -20:47:16.6 &   19.14 &  0.215 & 4b\\
& 509165 & 09:51:30.52 & -20:46:35.4 &   18.76 &  0.238 & 4c\\
& 518715 & 09:51:30.68 & -20:45:02.8 &   18.95 &  0.187 & 4a\\
& 489725 & 09:51:30.81 & -20:48:47.1 &   18.72 &  0.182 & 4a\\
& 510378 & 09:51:31.01 & -20:46:18.4 &   20.43 &  0.486 & \\
& 507808 & 09:51:31.34 & -20:46:41.6 &   19.61 &  0.236 & 4c\\
& 506364 & 09:51:31.45 & -20:46:54.1 &   17.05 &  0.236 & 4c\\
& 519886 & 09:51:32.71 & -20:44:44.2 &   20.19 &  0.214 & 4b\\
\hline
EIS0955-2113 & 325426 & 09:55:24.52 & -21:14:20.6 &   19.96 &  0.341 & 5a\\
& 326951 & 09:55:25.97 & -21:13:59.7 &   18.81 &  0.331 & 5a\\
& 328968 & 09:55:31.58 & -21:13:48.8 &   19.46 &  0.673 & 5b\\
& 328298 & 09:55:32.46 & -21:13:51.9 &   19.95 &  0.656 & 5b\\
& 329220 & 09:55:33.70 & -21:13:49.2 &   19.93 &  0.674 & 5b\\
& 323973 & 09:55:34.45 & -21:14:30.2 &   19.56 &  0.375 & \\
& 325779 & 09:55:40.01 & -21:14:10.2 &   20.09 &  0.279 & \\
\hline
\end{tabular}
\end{table*}

\addtocounter{table}{-1}
\begin{table*}
\caption[ ]{Continued.}
\label{t-lg}
\begin{tabular}{rrrrrll}
\hline
Field id. & Galaxy id. & $\alpha_{J2000}$ & $\delta_{J2000}$ & $m_I$ & $z$ &
galaxy set \\
\hline
EIS0956-2009 & 746450 & 09:56:20.17 & -20:09:35.9 &   20.73 &  0.349 & 6a\\
& 748248 & 09:56:22.38 & -20:09:33.8 &   19.88 &  0.443 & 6c\\
& 751377 & 09:56:24.32 & -20:09:09.6 &   19.00 &  0.447 & 6c\\
& 749243 & 09:56:25.24 & -20:09:14.7 &   20.55 &  0.451 & 6c\\
& 753860 & 09:56:26.21 & -20:08:32.8 &   19.76 &  0.443 & 6c\\
& 745962 & 09:56:26.33 & -20:09:41.4 &   20.26 &  0.362 & 6a\\
& 748889 & 09:56:27.66 & -20:09:03.0 &   18.26 &  0.363 & 6a\\
& 751960 & 09:56:28.11 & -20:09:01.7 &   19.36 &  0.445 & 6c\\
& 751382 & 09:56:28.96 & -20:09:05.2 &   18.80 &  0.444 & 6c\\
& 753734 & 09:56:29.04 & -20:08:40.0 &   19.63 &  0.369 & 6a\\
& 752280 & 09:56:31.79 & -20:08:53.9 &   19.38 &  0.402 & 6b\\
& 750180 & 09:56:34.08 & -20:08:46.0 &   19.49 &  0.392 & 6b\\
& 754800 & 09:56:35.18 & -20:08:28.0 &   19.00 &  0.441 & 6c\\  
\hline  		   		    	        	 
\end{tabular}
\end{table*}

\section{The definition of the galaxy systems in redshift space}
\label{s-zsys}

Since the six cluster candidates are drawn from the EIS cluster
catalogue, they obviously correspond to significant density enhancements
in projection. Here we search for systems of galaxy redshifts that could
be associated to the 2-d over-densities. In this way we assign a
spectroscopic redshift to 4 of our cluster candidates. We also
evaluate the probability that these systems correspond to a
genuine three-dimensional cluster.

Olsen et al. (\cite{ols99a}) search for clusters in projection
assuming a 2-d radial density profile with a cutoff radius
$r_{co}=1.33$ h$_{75}^{-1}$ Mpc. This size is well fitted to the
EFOSC2 field-of-view, corresponding to $1.9 \times 1.3$ h$_{75}^{-2}$
Mpc$^2$, at $z \sim 0.6$ (the average estimated redshift of our
candidate clusters). Therefore, we search for cluster members in
redshift-space within the whole EFOSC2 field.

Several refined algorithms for the definition of systems of galaxies in
redshift space can be found in the literature (e.g. Katgert et
al. \cite{kat96}; Pisani \cite{pis93}). However, with only a dozen
galaxy redshifts per field, these sophisticated algorithms can not be
applied. We choose to identify galaxy systems using a physical criterion based
on the well established properties of nearby clusters of galaxies.

Within each EFOSC2 field, we identify any
set of two or more galaxies contained within a given redshift range,
$\Delta z$.  In order to define $\Delta z$, we note that Abell-like
clusters of galaxies have mean velocity dispersions $\sigma_v \simeq
750$~km~s$^{-1}$ (Girardi et al. \cite{gir93}). Since the
line-of-sight velocity distributions of clusters are approximately
gaussian in shape (Girardi et al. \cite{gir93}), $>99$~\% of the
cluster members have a velocity within $\pm 3 \sigma_v$.  As a
consequence, galaxies in a given cluster should be located within a
redshift range $\Delta z \leq 0.015 \times (1+z)$, taking into account
the cosmological factor (Danese et al. \cite{dan80}).

We list in Table~\ref{t-cs} the sets of galaxies we identify in
redshift space. In Column (1) we list the EIS cluster field
identification, in Column (2) the galaxy set identification, in Column
(3) the number of galaxies in the set, in Column (4) the median
redshift of the set, in Column (5) the total redshift range covered by
the galaxies within the galaxy set. In Column (6) we list the
probability of the galaxy set to correspond to a significant
overdensity in redshift space, as estimated from resamplings of the
Canada-France Redshift Survey data-base (CFRS; Lilly et
al. \cite{lil95}; Le~F\`evre et al. \cite{lef95}; Hammer et
al. \cite{ham95}; Crampton et al. \cite{cra95}) -- see
Sect.~\ref{s-likely}. The real systems (probability $\geq 0.95$) are
flagged in Column (7).

\begin{table*}
\caption[ ]{Sets of galaxies in the EIS cluster fields}
\label{t-cs}
\begin{tabular}{rcrrrrc}
\hline
Field id. & galaxy & $N_{gal}$ & median$(z)$ & $\Delta z$ 
& galaxy set & $P \geq 95$~\% \\
& set & & & & probability & sets \\ 
\hline
EIS0533-2353 & 1a & 2 & 0.276 & 0.003 & 0.348   &  \\
          & 1b & 2 & 0.308 & 0.006 & 0.210   &  \\
          & 1c & 2 & 0.334 & 0.003 & 0.325   &  \\
          & 1d & 2 & 0.503 & 0.012 & 0.168   &  \\
EIS0540-2418 & 2a & 2 & 0.440 & 0.005 & 0.226   &  \\
          & 2b & 2 & 0.472 & 0.009 & 0.125   &  \\
          & 2c & 3 & 0.528 & 0.015 & 0.592   &  \\
          & 2d & 3 & 0.698 & 0.014 & 0.924   &  \\
EIS0950-2154 & 3a & 3 & 0.129 & 0.003 & 0.972   & Y\\
          & 3b & 2 & 0.214 & 0.001 & 0.632   &  \\
          & 3c & 2 & 0.241 & 0.002 & 0.500   &  \\
EIS0951-2047 & 4a & 2 & 0.184 & 0.005 & 0.478   &  \\
          & 4b & 2 & 0.214 & 0.001 & 0.632   &  \\
          & 4c & 5 & 0.236 & 0.007 & 0.993   & Y\\
EIS0955-2113 & 5a & 2 & 0.336 & 0.010 & 0.326   &  \\
          & 5b & 3 & 0.673 & 0.018 & 0.957   & Y\\
EIS0956-2009 & 6a & 4 & 0.362 & 0.020 & 0.714   &  \\
          & 6b & 2 & 0.397 & 0.010 & 0.063   &  \\
          & 6c & 7 & 0.445 & 0.010 & 0.997   & Y\\
\hline\end{tabular}
\end{table*}

In total we identify 19 galaxy sets along the line-of-sight of
six EIS candidate clusters. It is interesting to detail the comparison of
 the estimated
mean redshifts, $z_{mf}$'s, of these clusters (see Table~\ref{t-sc}),
to the spectroscopic redshifts of the 19 sets (see
Table~\ref{t-cs}).  In this comparison, we take into account the
uncertainties in the mean redshifts of the galaxy sets, and note that the
matched-filter redshift estimates are at most accurate to within
$\Delta z = \pm 0.05$.

In the case of the candidate clusters EIS0533-2353, EIS0950-2154, and
EIS0951-2047, we do not find any galaxy set close to the estimated cluster
redshifts. According to the matched-filter algorithm, these three
clusters are located at a higher redshift than any of the galaxy
sets found in their fields. In each of the fields of EIS0955-2113 and of
EIS0956-2009, there is one set of galaxies with mean redshift
close to the estimated redshift ($\overline{z}=0.673$
vs. $z_{mf}=0.6$, and $\overline{z}=0.445$ vs. $z_{mf}=0.5$,
respectively).  Finally, in the field of EIS0540-2418, there are two
sets of galaxies with mean redshifts close to the cluster estimated
redshift, $z_{mf}=0.6$.

The detection of galaxy concentrations close to the estimated
redshifts of three EIS clusters supports the
reliability of the matched-filter redshift estimates.  

As far as the failed detections are concerned, EIS0533-2353 may have
escaped detection because of its high (estimated) redshift, $z_{mf} =
0.7$ -- we may simply have not been observing deep enough.  Moreover,
the line-of-sight to a single cluster can intercept several different
galaxy sets. Katgert et al. (\cite{kat96}) estimate that 10~\% of nearby
Abell clusters result from the superposition of two almost equally
rich systems (and this fraction is probably higher for more distant
clusters). The nearest of these systems has the highest chance of
being detected.  In this context, we note that the low-$z$ set of galaxies,
3a, detected in the field of EIS0950-2154 is somewhat off-centered with
respect to the nominal EIS cluster center (see Fig.~\ref{f-m3}).
The same is not true for the sets with $z \sim z_{mf}$ in the fields
of EIS0540-2418, EIS0955-2113, and EIS0956-2009. This fact suggests that
the set 3a does not correspond to the EIS cluster, but is a foreground
group.

Da~Costa et al. (\cite{dac00}) suggest that all the six EIS candidate
clusters could be real clusters at redshift $> 0.5$. Da~Costa et
al. base their suggestion on the detection of red-sequences of
early-type galaxies in the colour-magnitude diagrams. They note that
under-sampling of the redshift distribution of galaxies in the cluster
fields may explain the lack of spectroscopic detections of some clusters.

\section{The reality of the galaxy sets}\label{s-likely}
\begin{figure}
\resizebox{\hsize}{!}{\includegraphics{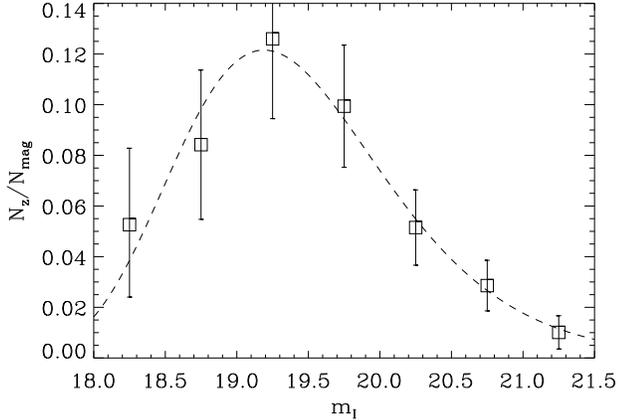}}
\caption{The function $s(m)$, computed as the ratio
between the binned magnitude distributions of galaxies with redshifts,
and of galaxies with magnitudes, in the range $17 \leq m_I \leq 21.3$,
in the six EIS cluster fields. The dashed line is the best fit to the data
with a Schechter function.}
\label{f-sf}
\end{figure}

In this Sect. we assign likelihoods to the 19 sets of galaxies which
have been identified in the previous Sect.  If $s(m)$ is the
incompleteness of our redshift sample in the magnitude interval
$[m_1,m_2]$ (see Fig.~\ref{f-sf}), then the redshift distribution
reads
\begin{equation}\label{e-dndz}
N(z) dz= \frac{dV(z)}{dz} dz \int_{m_1}^{m_2} f[L(m,z)] \; s(m) \; dm,
\end{equation}
where $V(z)$ is the volume element at the redshift $z$ and $f[L(m,z)]$
is the luminosity function (LF hereafter).  We take the LF
in the $I_4$ band given by Postman et al. (\cite{pos96}) and convert
it to our $I_c$ band, using the Postman et al. transformation
procedure. Our reference LF has therefore a Schechter (\cite{sch76})
form with parameters $\alpha=-1.1$ and $M^{\ast}=-22.15+ 5 \log
h_{75}$. We also assume negligible evolution of the LF out
to $z \sim 0.8$ (see Lilly et al. \cite{lil95} and Lin et
al. \cite{lin99}). Our LF then depends on $z$ only through the
evolution of the stellar populations and the K-correction, that we
take from Poggianti (\cite{pog97}).

\begin{figure}
\resizebox{\hsize}{!}{\includegraphics{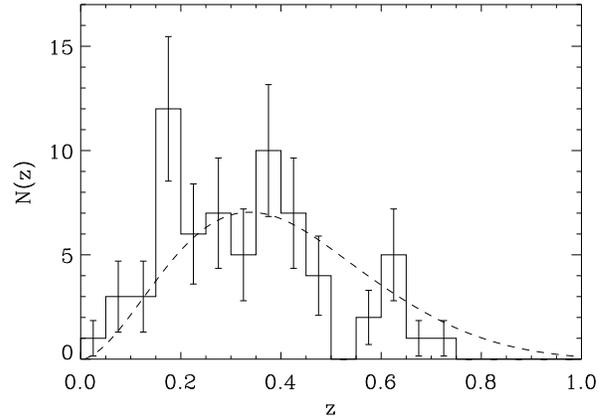}}
\caption{The redshift distribution function, $N(z)$ (dashed line) and
the redshift histogram of the 67 galaxies of our sample in the
redshift range $0.0 \leq z \leq 0.8$. $N(z)$ is normalised to the same
number of objects, 67.}
\label{f-nz}
\end{figure}

In Fig.~\ref{f-nz}, we plot $N(z)$, computed according to Eq.~1, along
with the observed $z$-distribution. Based on the estimated $N(z)$, we
find that 16 out of 19 sets correspond to significant overdensities in
redshift space, with a probability $\geq 95$~\%. Since Eq.~1 provides
$N(z)$ for a uniform galaxy distribution, these probabilities do not
include the effect of large scale clustering (see, e.g., Zaritsky et
al. \cite{zar97}).  In order to account for the large scale
clustering, and following the approach by Holden et al. (\cite{hol99a}),
we modulate $N(z)$ with the redshift distribution derived from the
CFRS assuming our selection function, $s(m)$.

We first convert the CFRS $I_{AB}$ magnitudes to the $I_c$ system (see
Lilly et al. \cite{lil95}). We then extract 50000 galaxies from the
CFRS, with a bootstrap sampling, adopting the magnitude distribution
of our sample (see Fig.~\ref{f-md}).  A Kolmogorov-Smirnov test
shows that the bootstrapped CFRS and our data-set have similar
redshift distributions. This is expected since magnitude selection is the
main process leading to the inclusion of galaxies in the sample.

We extract random subsamples of 11 galaxies from the bootstrapped CFRS
reference sample (11 is the mean number of galaxies with redshift in
our EIS cluster fields). Using the same procedure described in
Sect.~\ref{s-zsys}, we identify sets of galaxies within these subsamples,
and compute their probabilities relative to the uniform redshift
distribution $N(z)$ given by Eq.~1. In this way we construct a
distribution of probabilities to detect a real system within a galaxy
sample which includes large-scale clustering, but not galaxy
clusters. We finally obtain the likelihoods of the 19 observed sets,
by comparing their original $N(z)$-based probabilities to the
distribution of probabilities for the random sets. These 19 likelihoods
are listed in Table~\ref{t-cs}.

We find that four of our 19 sets have a likelihood $> 95$~\%; all of
them have at least three galaxy members.  The four sets are flagged in
the last column of Table~\ref{t-cs}. We refer to these four sets as
the 'real systems' hereafter.  As expected, many of the sets with a
significant overdensity with respect to the uniform redshift
distribution are no longer significant when compared to a redshift
distribution which includes the large scale clustering.

Our results are robust against modifications of the adopted LF (we
change $\alpha$ by $\pm 0.2$, and $M^{\ast}$ by $\pm 0.5$ mag), and of
the galaxy-type for which the evolutionary- and K-corrections are
computed. Furthermore, we verify that varying cosmological parameters
(h$_{75}$, $\Omega_0$ and $\Omega_{\Lambda}$) within conservative
ranges, only induces marginal changes in the system
likelihoods. Finally, we also checked that narrowing the $\Delta z$
range used to define the redshift-sets, from 0.015 to 0.010, hardly
modifies the membership and likelihoods of the sets.

\section{Discussion and conclusions}\label{s-con}

We obtain 67 new redshifts for galaxies in six EIS candidate cluster
fields. Based on these data, we establish the existence of real
systems in redshift space in the direction of four of these
candidate clusters. The reality of the systems is established at $>
95$~\% confidence level, and in two cases, at $> 99$~\%. The redshift
overdensities, coupled with the 2-d overdensities detected by the use
of the matched-filter algorithm, strongly supports the reality of four of the
six examined EIS clusters. These 4 clusters add to the other
two spectroscopically confirmed EIS clusters (da~Costa et al. \cite{dac99}).

Two of the four $z$-systems have a median redshift in good agreement
with the matched-filter estimate for the EIS cluster redshift ($\mid$
median$(z)-z_{mf} \mid < 0.1$). The other two have significantly lower
redshifts (median$(z) = 0.129$, 0.236 vs. $z_{mf}=0.5$).  

Taken at face value, these results suggest that, in several cases, the
matched-filter algorithm over-estimates the mean cluster redshift by a
large amount. However, it is quite possible that in some cases we have
not detected the EIS cluster, but a foreground galaxy system projected
along the same line-of-sight of the cluster. Similarly, it is
difficult to conclude about the reality of the EIS clusters where we
do not detect any real redshift system.

In particular, we note that in the field of EIS0540-2418 we have a
marginal detection (92~\% probability) of the galaxy set 2d at
median$(z)=0.698$ (see Table~\ref{t-cs}), in fair agreement with the
matched-filter estimate of the cluster redshift,
$z_{mf}=0.6$. We also note that the cluster EIS0533-2353 has
$z_{mf}=0.7$, larger than any of our galaxy sets. This suggests that
it could have escaped detection because our observations were not deep
enough. In fact, da~Costa et al. (\cite{dac00}) suggest that all our
six EIS cluster candidates could be real, based on the analysis of the
colour-magnitude diagrams for galaxies in the cluster fields.

We conclude that our spectroscopic confirmation rate must be
considered as a lower limit. If at least one third of the EIS clusters
in the redshift range sampled by our observations are real, there are
more than 25 EIS clusters with $z_{mf}$ in the range 0.5--0.7.  This
sample is large enough for the derivation of the properties of
clusters at intermediate to high redshifts.

Optical selection of clusters of galaxies at high redshifts is a
necessary complementary approach to X-ray selection. While X-ray
selection tends to detect only rich Abell-like clusters, optically
selected cluster samples contain a large number of poor clusters. In
fact, the space density of PDCS clusters is five times higher than
that of rich Abell clusters (Holden et al. \cite{hol99a}), and very
few PDCS clusters are X-ray bright (Holden et al.  \cite{hol97}).
Consistently, the velocity dispersions of our two systems with $\geq
5$ galaxy redshifts (system 4c, at $\overline{z}=0.236$ and system 6c,
at $\overline{z}=0.445$, see Table~\ref{t-cs}) are $\sim
600$~km~s$^{-1}$, typical of low-richness clusters ($R \leq 1$, see
Girardi et al. \cite{gir93}).

With the current and near-future ground-based facilities for
wide-field optical and near-infrared imaging, we can expect a rapid
increase in the samples of optically selected clusters. Currently, our
spectroscopic sample only comprises $\sim 10$~\% of all the clusters
in the two patches C and D, and in the (estimated) redshift range
0.5--0.7. We plan to extend our sample in forthcoming observing runs.
Confirmed EIS clusters at high redshift will be the natural targets
of VLT observations aimed at determining their dynamical properties.

\begin{acknowledgements}
We thank Nando Patat for imaging the six EIS fields in preparation for
our spectroscopic run. 
\end{acknowledgements}

\end{document}